\documentclass [11pt,a4paper,amsmath, aps, amssymb, bbm, nofootinbib, noindent, notitlepage] {article}

\usepackage[british]{babel}

\usepackage[a4paper,top=2cm,bottom=2cm,left=2.5cm,right=2.5cm,marginparwidth=1.75cm]{geometry}

\usepackage[backend=biber, style=apa, eprint=true]{biblatex}
\usepackage{xpatch}
\xpatchbibdriver{article}
  {\usebibmacro{doi+url}}
  {\usebibmacro{doi+url}%
   \newunit\newblock
   \usebibmacro{eprint}}
  {}
  {}

\usepackage{amsmath}
\usepackage{graphicx}
\usepackage[colorlinks=true, allcolors=blue]{hyperref}
\usepackage{hyperref}
\usepackage{orcidlink}
\usepackage[title]{appendix}
\usepackage{mathrsfs}
\usepackage{amsfonts}
\usepackage{booktabs} 
\usepackage{caption}  
\usepackage{threeparttable} 
\usepackage{algorithm}
\usepackage{algorithmicx}
\usepackage{algpseudocode}
\usepackage{listings}
\usepackage{enumitem}
\usepackage{chngcntr}
\usepackage{booktabs}
\usepackage{lipsum}
\usepackage{subcaption}
\usepackage{authblk}
\usepackage[T1]{fontenc}    
\usepackage{csquotes}       
\usepackage{diagbox}

\usepackage{setspace}
\usepackage{titlesec}
\titleformat{\section} 
  {\normalfont\Large\bfseries}{\thesection.}{1em}{}
  
\usepackage{lineno} 

\rightlinenumbers 

\usepackage{float}   
\usepackage{caption} 


\usepackage{amsmath,amsfonts,amssymb,bbm, mathtools, mathrsfs}
\usepackage{hyperref, csquotes}
\usepackage{graphicx, color, svg,subcaption}
\usepackage{tikz,pgf,tikz-cd,float}
\usepackage{physics,tensor}
\usepackage{rotating}
\include{fdiagrams}
\definecolor{timelike}{RGB}{227, 11, 91}
\definecolor{spacelike}{RGB}{0, 128, 128}
\definecolor{lightlike}{RGB}{0, 25, 150}

\title{The quantum gravity seeds for laws of nature}

\author[1,2]{Vincent Lam}
\author[3, 4]{Daniele Oriti}
\affil[1]{\small Institute of Philosophy, University of Bern, CH-3012 Bern, Switzerland}
\affil[2]{\small School of Historical and Philosophical Inquiry, The University of Queensland, St Lucia, QLD 4072, Australia}
\affil[3]{Depto. de Física Teórica, Facultad de Ciencias Físicas, \\ Universidad Complutense de Madrid Plaza de las Ciencias 1, 28040 Madrid, Spain, EU}
\affil[4]{Munich Center for Mathematical Philosophy, Ludwig-Maximilians-Universit\"at M\"unchen, Ludwigstrasse 31, 80333 M\"unchen, Germany, EU}

\date{}  

\begin{document}
\maketitle

\begin{abstract}
We discuss the challenges that the standard (Humean and non-Humean) accounts of laws face within the framework of quantum gravity where space and time may not be fundamental. This paper identifies core (meta)physical features that cut across a number of quantum gravity approaches and formalisms and that provide seeds for articulating updated conceptions that could account for QG laws not involving any spatio-temporal notions. To this aim, we will in particular highlight the constitutive roles of quantum entanglement, quantum transition amplitudes and quantum causal histories. These features also stress the fruitful overlap between quantum gravity and quantum information theory.    

\end{abstract}


\section{Introduction}\label{introduction}

What laws of nature are is at the same time a central issue in philosophy and a question that lies at the heart of science. In a naturalistic perspective, the expectation is that the latter informs the former, so that the philosophical understanding of laws can account for those `discovered' by science. Indeed, the venerable philosophical (metaphysical) debate about the nature of laws has been enriched in recent decades by considerations from science, and in particular from fundamental physics. As a crucial example, quantum entanglement (underlying Bell-type non-local correlations) has been famously argued to be in tension with the Humean account of laws, and more specifically with the thesis of Humean supervenience (see e.g. \cite[ch. 2]{Maudlin2007}). In the light of this tension, proposals have emerged to refine the Humean conception, for instance by incorporating fundamental quantum entanglement relations in the Humean supervenience basis (see e.g. \cite{Darby2012}), i.e. as constitutive and `gluing' elements of the Humean mosaic (alongside the spatio-temporal relations). 

Now, quantum gravity (QG) research (\cite{Seiberg,Padmanabhan,OritiDisEmerg}) is suggesting a radical revision of our understanding of the physical world, according to which standard spatio-temporal features and spacetime itself may not be fundamental (for philosophical discussions, see \cite{Huggett-Wuthrichforth, OritiLevels,Crowther}), but only emergent in some (e.g. functionalist) sense (see \cite{lamwut18,MargoniOriti}). Such a radical revision––suggested to different degrees by most of the approaches and formalisms to quantum gravity––would have a drastic impact on our philosophical accounts of what laws of nature are, since the two main families of standard conceptions about laws––on the one hand: Humean, `descriptive' conceptions; on the other hand: non-Humean, `production' or `governing' conception––depend on spacetime in one way or another (\cite{Lam:2023aa}). Surprisingly, very little work has been done so far to amend the standard philosophical account of laws in the light of this radical suggestion from the research in quantum gravity––beside \cite{Lam:2023aa}. We build on this latter work, which mainly diagnoses the problems with the current accounts, and aim to identify fundamental QG features that are common to a number of QG approaches and formalisms and that could help to articulate the standard accounts of laws in a context where spacetime may be absent altogether.  

For each of the main (family of) standard conception(s) about laws, \cite{Lam:2023aa} have identified major obstacles when trying to amend them in a way that would avoid their reliance on spacetime. In the absence of the latter, a Humean account needs a categorical (non-modal, non-nomic) relation that holds the mosaic of facts together. \cite{Lam:2023aa} hint at quantum entanglement (underlying Bell-type correlations) as a possible gluing relation in QG (see also \cite{Jaksland:2021aa}), without grounding in detail this proposal in the actual QG approaches and formalisms though. When it comes to the non-Humean conceptions––such as primitivism or dispositionalism about laws––the crucial challenge is about articulating a notion of non-temporal production (\cite[\S 3.2]{Lam:2023aa}); however, what this latter could look like in QG is mainly left open.

This paper aims to remedy these gaps by identifying core (meta)physical features that cut across a variety of QG approaches and formalisms (a relevant endeavour in itself) and that would contribute to the articulation of revised conceptions of laws, which would also account for those to be discovered at the QG level––something that is currently not the case. 

Before moving on, a few comments are in order. First, we will not delve into the metaphysical intricacies that are related to the detailed articulation of the various accounts of laws (including their revised versions). The scope of this paper is restricted to the (limited, but crucial) task of specifying the physical QG features that could constitute a basis for revised (or novel) accounts of laws. 

Second, there is no accepted theory of quantum gravity as of yet, nor experimentally corroborated aspects of any of the current proposals, except of course for their starting points, i.e. classical general relativity (GR) for the gravitational interaction (and spacetime) and quantum field theory (QFT) for matter and non-gravitational interactions, which they have to reproduce in the appropriate regime. There is instead a diversity of proposed approaches and formalisms (see \cite{OritiApproaches,Handbook}), many exciting partial results and conceptual insights. We will not do justice to such array of results or approaches, and limit ourselves to take onboard some suggestions which are somehow shared across several quantum gravity formalisms and have, we believe, foundational significance––in particular, beyond the interpretative issues related to quantum theory. In this perspective, we will start with the radical revisions of the classical notions of space and time that are suggested by most of the QG approaches and the challenges they raise for the main standard conceptions of laws (sections \ref{disappearance} and \ref{structures}). Then we will highlight the constitutive role played by quantum entanglement in these approaches, thus providing the QG seeds (glue) of the Humean mosaic (section \ref{entanglement}). Similarly, in section \ref{transition}, we will argue that non-temporal notions of dynamics and (proto-)causality in terms of transition amplitudes constitute a central piece of many QG formalisms, and could provide non-Humean accounts of laws with a physically meaningful notion of non-temporal production (or governance). 

Third, faced with the possibility that spacetime is not fundamental but emergent and concerned with the ensuing impact on the concept of laws of nature, one could take a straightforward if radical move: if the notion of laws does indeed require spacetime and if the latter is emergent, then so are laws. Thus there are simply no laws at the fundamental (QG) level: there are no `laws of quantum gravity'. This is logically straightforward, and we have no in-principle objections against this position. It then leaves us with the task of understanding physics––or at least: quantum gravity––without calling for the notion of laws at all. This is an interesting task, but one we leave for another time. Here, we rather ask what notion of law can survive the disappearance of spacetime, and still apply in a quantum gravity context.    

Finally, and in the same line of thought as the previous comment, there are two main avenues for characterizing laws of nature: an ontic perspective, based on the idea that laws are `out there' in the natural world, and an epistemic perspective, in which laws are agents' constructions and thus exist only in an epistemic realm––skipping over a number of (important) subtleties. We will mostly deal with the first kind of view, which include the standard Humean and non-Humean conceptions mentioned above, because they are more radically affected by the spacetime disappearance, but we will also comment on the second kind of view, which is affected too albeit in a less direct manner (section \ref{epsitemic}).  

\section{Disappearance of spacetime in quantum gravity and the challenges to laws}\label{disappearance}
Our common-sensical understanding of space and time, often naively imported in philosophical considerations, is challenged in many ways already by classical general relativity. The theory teaches us that there are no preferred spatial or temporal directions, that spacetime geometry is generically non-flat and, most challenging for our common sense, dynamical and dependent on matter configurations. Moreover, so is the causal structure, i.e. the set of (possible) cause-effect relations among spacetime-localized systems. Furthermore, although general relativity, like the rest of classical physics, is formulated in terms of smooth fields living on a differentiable manifold, often naively understood as the set of spacetime events, the diffeomorphism invariance of the theory challenges this naive identification, and deprive manifold points and directions of physical significance, requiring us (in the most accepted strategy) to use fields themselves to define events and spacetime localization in a physically meaningful manner. In this `relational' strategy, some physical degrees of freedom (e.g. matter fields) are used to define a physical reference frame (clock and rods) in order to specify the localization and evolution of other physical degrees of freedom. 


When moving to the quantum domain, things get `worse' for space and time––and in a sense, also more interesting! At a bare minimum, maintaining as fundamental the same kind of entities on which GR is based, i.e. metric (geometry) and matter fields, their acquiring quantum properties already deeply shatters the general relativistic spatio-temporal picture based on relational localization and dynamical geometry. Geometric and spacetime-characterized observables will, in general, fail to commute, thus making localization, causality and geometry itself only `approximately determinable'––of course, what this exactly means will also depend on the way the quantum interpretative issues are addressed (what is crucial to see here is that this `quantum indeterminacy', however its precise interpretation, concerns the metric field itself as well as the fields used to define physical reference frames). There will be no quantum state of fields (and geometry) that will encode precisely all spacetime properties (again, whatever the interpretation)––the latter may only acquire unique values in a classical approximation. Entanglement and quantum contextuality render spacetime properties ever harder to identify: this follows from the expectation that the GR fields themselves will become quantum systems. While details will depend on specific quantum gravity frameworks (as well as on the exact interpretation of the quantum formalism), this general picture is independent of whether one uses a canonical or path integral quantization, or any other quantization method.

The original assumption that the fundamental theory is going to be based on quantum fields and obtained from some quantization of general relativity is, however, likely to be too restrictive. The more fundamental theory may be built on new quantum degrees of freedom, new physical entities possibly described by new mathematical structures, with geometry and matter as smooth fields arising as collective, effective, emergent quantities only in some approximation. This is an hypothesis supported by a number of results coming from semiclassical physics as well as modern quantum gravity formalisms (\cite{Seiberg,Padmanabhan,OritiDisEmerg}). The precise kind of new structures replacing spatiotemporal fields varies in different quantum gravity formalisms (\cite{OritiApproaches,Handbook}). What is shared is however the challenge of reconstructing spacetime, in some approximation, from the collective dynamics of the new quantum entities. This is a new challenge that is distinct, both technically and conceptually, from the issue of semiclassical approximation (and the issues related to it in quantum foundations, e.g. the quantum measurement problem). 



The `disappearance of spacetime' in the QG context––to different degrees depending on the approach––raises straightforward difficulties for the main standard conceptions of laws, to the extent that they reply on spacetime. We will now briefly recall these challenges for the main Humean and non-Humean accounts (we refer to \cite{Lam:2023aa} for a detailed discussion). Within the broad Humean framework, the general ontological picture is that of a fundamental \emph{spatio-temporal} basis of non-modal facts, referred to as the `Humean mosaic', over which laws supervene (see \cite{Hall2015} for recent discussion). The Humean mosaic can be characterized as a distribution of fundamental intrinsic properties (“perfectly natural properties” in Lewis' terms, see \cite[x]{Lewis1986}), absent of any necessary connections among such properties, solely connected by \emph{spatio-temporal} relations holding them together (and thus providing some non-modal `glue'). The most developed incarnation of a Humean account of laws is the so-called `best system' one (see e.g. \cite{Earman1984}) (or its most recent version, the `better best system', \cite{Cohen:2009aa}), according to which, laws are theorems (logically deduced consequences) in our best theoretical systems describing the natural world, where theoretical systems are evaluated on the basis of descriptive power and simplicity and are grounded on the Humean mosaic of \emph{spatio-temporal} relations among categorical facts. In this Humean framework, and as we mentioned in section \ref{introduction}, the disappearance of spacetime in QG raises the question of what plays the role of the non-modal glue holding the Humean mosaic together. From a Humean perspective, there seems to be no clear ontic basis for QG laws anymore. Within a best system characterization of laws, one could simply maintain that QG laws will still be the theorems in our QG best system, i.e. the final, agreed-upon, theory of QG. However, the issue will then be that neither the underlying Humean mosaic nor (most of) the theorems in such QG best system would refer to spacetime. The fundamental question that arises in this context concerns the ontological characterization of the non-spatio-temporal Humean mosaic.

Similarly, within the non-Humean framework, the primitivist and dispositionalist conceptions of laws face major challenges in the QG context without spacetime. The general ontological picture underlying these conceptions relies on some irreducible, primitive modality which gives rise to the spatiotemporal distribution of particular facts. According to primitivism, this irreducible modality arises from fundamental physical laws taken as ontological primitives (\cite{Maudlin2007}), not grounded on anything else. Dispositionalism considers that laws are in fact grounded in the fundamental and primitive dispositional (or causal) nature of properties (\cite{Bird2007}), which also gives rise to some irreducible modality. What is crucial to stress here is that according to both primitivism and dispositionalism about laws, it is a proper subset of particular fundamental facts endowed with some primtive modality that (ontologically) grounds or (nomically) \emph{produces} the entirety of the distribution of facts. In fact, these accounts of laws actually relies on a \emph{temporal} notion of nomic production (according to both primitivism and dispositionalism, laws operate or govern against a temporal background, see \cite[\S 3]{Lam:2023aa}). In the non-(spatio-)tempral setting of QG, the fundamental question for these non-Humean conceptions is then how to generalize the notions of nomic production and governance such that they do not rely on (space)time.

Before addressing these challenges, let us first specify the kind of new QG structures replacing spatiotemporal fields in some important QG approaches.

\section{Examples of quantum gravity structures}\label{structures}

In canonical loop quantum gravity (LQG) (\cite{LQG}), spin foam models (\cite{SF}), tensorial group field theories (in the more quantum geometric models, usually indicated as group field theories, see \cite{GFT}) and in fact in lattice quantum gravity based on connection-tetrad variables, quantum space and geometry are to be reconstructed from spin networks. These are graphs labeled by algebraic data taken (for 4d models in Lorentzian signature) from the representation theory of the Lorentz group $SL(2,\mathbb{C})$ or its rotation $SU(2)$ subgroup (depending on the specific model).

\begin{figure}[!ht]
\centering
\includegraphics[width=0.3\linewidth]{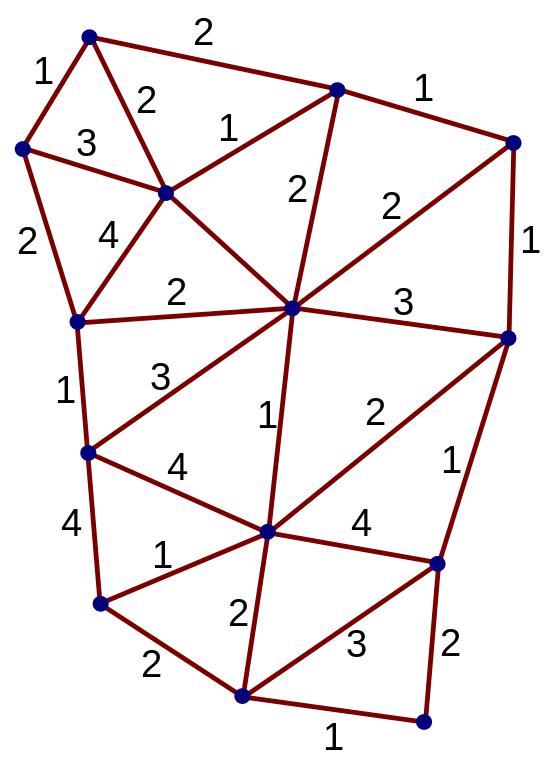}
\caption{\label{fig:frog}An example of spin network, labelled by irreps of $SU(2)$, i.e. spin labels (picture taken from Wikipedia).}
\end{figure}

For a given graph, spin network states, corresponding to all possible assignments of group irreducible representations (irreps) to the links of the graph and intertwiners between irreps at vertices, form a complete basis for the Hilbert space $\mathcal{H}_\gamma$ of the fundamental degrees of freedom of quantum space. One has then to consider the Hilbert space including states associated to all possible graphs. The precise construction is different in different formalisms. In canonical LQG the graphs are understood as embedded in an auxiliary spatial manifold; this results in the spin network states carrying additional topological and knotting information, being subject to diffeomorphism transformations––so that only equivalence classes of spin networks under diffeomorphisms have physical meaning––and in additional equivalence relations being imposed on states associated to different graphs (cylindrical equivalence), resulting from properties of the continuum connection field one wishes to quantize. The result is a Hilbert space including infinitely refined graphs and incorporating much of what a `quantized gravitational connection' would correspond to: 

\begin{equation}
\mathcal{H} = lim_{\gamma} \frac{\bigcup_\gamma \mathcal{H}_\gamma}{\approx} = L^2(\bar{\mathcal{A}})
\label{eq1}
\end{equation}
where we have schematically indicated the limit of infinitely refined graphs and the equivalence relations among them, turning the union of Hilbert spaces for different graphs into a proper Hilbert space, corresponding to (square integrable) functions of a generalised connection field. Despite the additional requirements, one ends up with a theory of distributional connections fields and geometries, thus requiring much additional work to recover the smooth fields of classical GR.

In other contexts, e.g. some spin foam model constructions, one drops the embedding information as well as the equivalence relations between spin networks on different graphs, and considers more abstract spin network states, and a Hilbert space of the form $\mathcal{H}^{'}= \bigoplus_{\gamma} \mathcal{H}_\gamma$. This also means that spatial topology is reconstructed from the combinatorial structure of the spin network graphs. This is the case for spin networks and spin foam constructions considered from a lattice quantum gravity perspective. Here spin network vertices (with associated link data) are taken to correspond to quantized 3-simplices, with spin labels carrying spectral properties of discrete geometric operators (triangle areas):

\begin{figure}[!ht]
\centering
\includegraphics[width=0.3\linewidth]{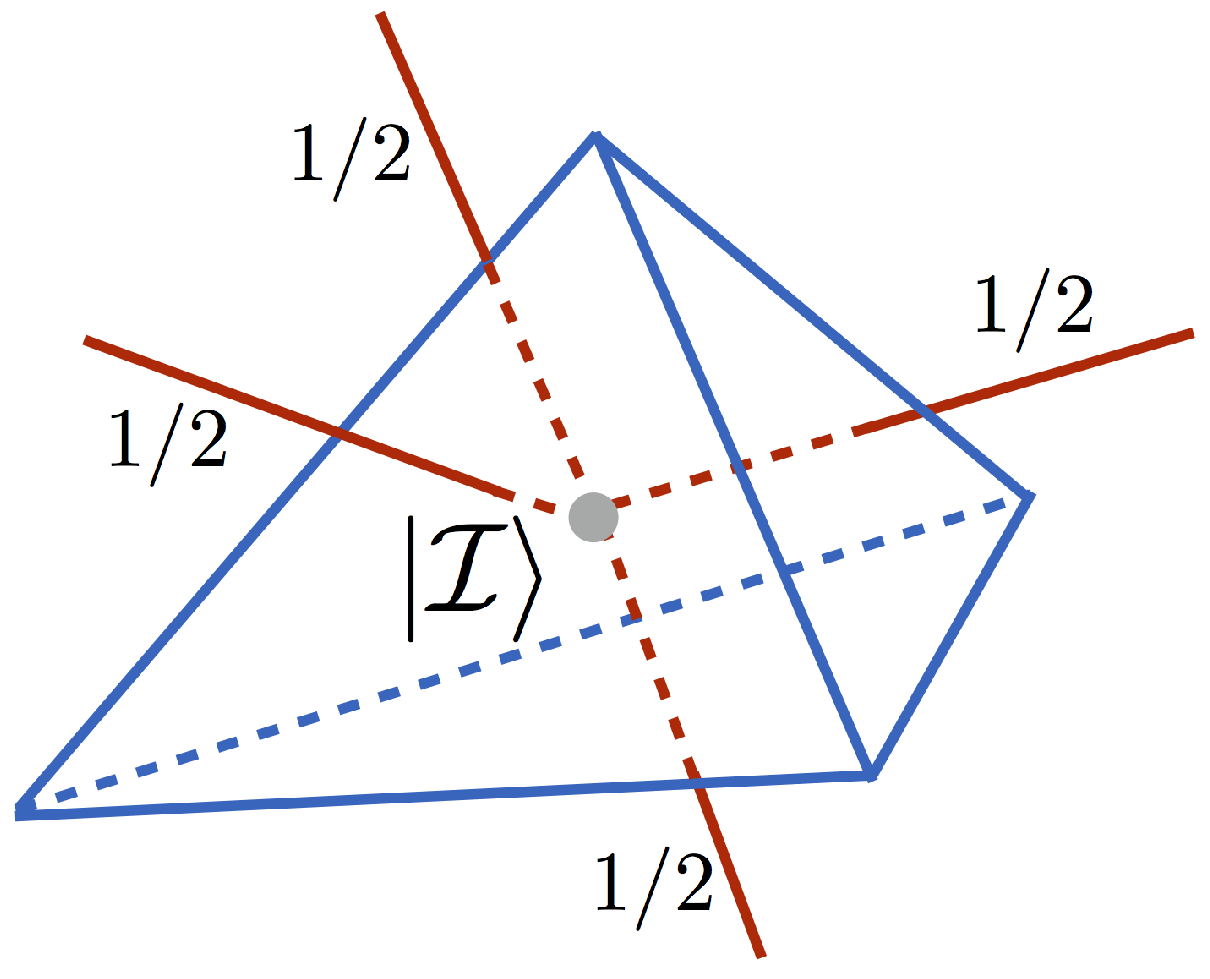}
\caption{A spin network vertex, with specific choice of labels, dual to a 3-simplex (tetrahedron).}
\end{figure}

Spin network states can then at best reproduce piecewise-flat simplicial geometries in a semi-classical approximation.  

In spin foam models, and even more decisively in the tensorial group field theory formulation, one can take an even more abstract perspective on spin networks––more detached from any continuum geometric interpretation––and take them as quantum many-body states made out of their single-vertex building blocks, i.e. quantum tetrahedra (for 4d models), each with a Hilbert space $\mathcal{H}_ v$. The corresponding Hilbert space can be taken to be a Fock space of such abstract quanta
\begin{equation}
\mathcal{F}(\mathcal{H_v}) = \oplus_{V=0}^\infty sym \left\{ \left( \mathcal{H}_v^{(1)}\otimes \cdots \otimes \mathcal{H}_v^{(V)}\right)\right\} \equiv \oplus_{V=0}^\infty sym \left\{ \mathcal{H}_V\right\}
\label{eq2}
\end{equation}
so that the Hilbert space of spin network states associated to a graph with $V$ vertices is contained in the Hilbert space of $V$ space quanta $\mathcal{H}_V$. We will say more about how the gluing of vertices/tetrahedra to form extended graphs/complexes is enforced in the following. At this QG level, we have thus a picture of a (non-spatio-temporal) many-body system (sometimes referred to as `quantum space' in the QG literature), with generic quantum states being far away from any charaterization (and mathematical encoding) in terms of continuum spatiotemporal fields.

When considering the quantum dynamics, the same combinatorial and algebraic `replacement' of spatiotemporal fields is implemented in the mentioned quantum gravity formalisms. 

The dynamical aspect to be imposed on quantum gravity states would have to be encoded in a specific set of constraint equations, in the form of Schwinger-Dyson equations for quantum correlation functions in a spin foam or tensorial group field theory context, or Hamiltonian constraint in a canonical loop quantum gravity setting. Such constraint equations and the observables onto which one would impose them would have no temporal dependence, in the form of absolute temporal parameters labelling evolution processes. Moreover, other complications (with respect to intuitive and classical notions of evolution) may arise; for example: quantum initial value problems may not be well-posed, unitarity may not be an exact requirement on quantum dynamics (for lack of absolute temporal evolution), topology may not be fixed but instead dynamical itself.

Again, there are several differences among the different quantum gravity formalisms mentioned above, in the precise encoding and formulation of the quantum dynamics––in fact there are several ways of formulating it even within the same approach. However, in all of them, `histories' (in a generalized, non-temporal sense) of the fundamental quantum entities---the possible elementary `processes' (again, in a non-temporal sense)---correspond to spin foams, i.e. 2-complexes labelled by the same algebraic data labelling spin networks, sometimes augmented to provide cellular complexes with the same data. 

\begin{figure}[!ht]
\centering
\includegraphics[width=0.8\linewidth]{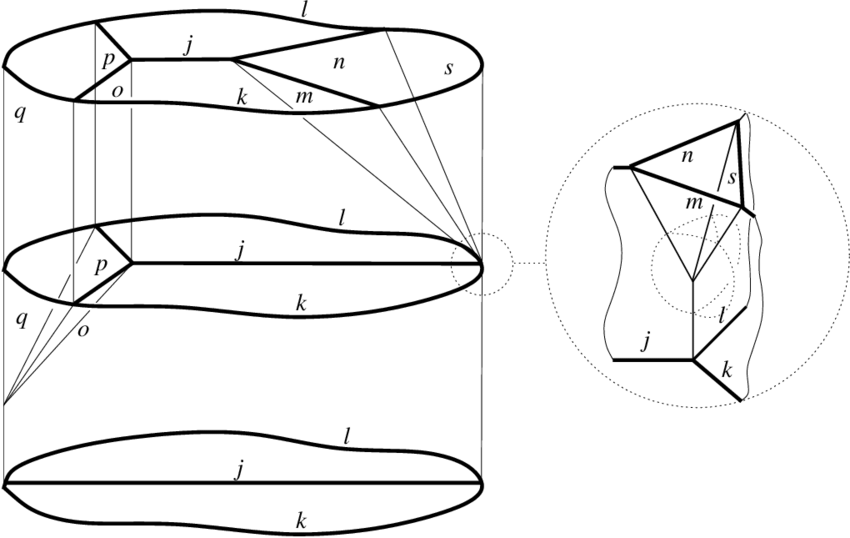}
\caption{A spin foam representing a possible \lq history \rq of (3-valent) spin networks (with specific choice of labels) - courtesy of A. Perez.}
\end{figure}

A `transition amplitude' between spin network states will then be given by a sum over spin foams, generically involving both a sum over complexes (bounded by the spin network graphs) and a sum over associated algebraic data (with fixed algebraic data on the boundary). This can be seen as a combinatorial and algebraic counterpart of a gravitational path integral, and indeed, for each given complex and appropriate spin foam amplitudes, it can be directly related to a lattice gravity path integral on the (simplicial) complex dual to the spin foam:

\begin{equation}
\langle \Phi_\gamma(j,i) | \Psi_{\gamma'}(j',i') \rangle = \sum_{\Gamma | \gamma, \gamma'} w(\Gamma) \sum_{\{J, I\} | j,j',i,i'} \mathcal{A}_\Gamma\left( J,I\right) \, \approx\, \int \,\mathcal{D}g \; e^{i S(g)} \; .
\label{spin foam model}
\end{equation}
In the tensorial group field theory context, this expression for the transition amplitudes of spin network states is obtained in the perturbative expansion of the field theory for their constitutive quanta of `space'.

\

Beyond the technical details of (and differences between) the quantum gravity formalisms sharing the above structures, there is a widespread consensus that the latter are hard to interpret in spatio-temporal terms––hence the claims about the `disappearance of spacetime' at the quantum gravity level. But then many questions arise about what that exactly means and implies. How should be understood the (meta)physical, but non-spatio-temporal picture suggested in these QG approaches? In such a non-spatio-temporal context, what grounds, \emph{what constitutes the seeds for quantum gravity laws}? More specifically, what holds the Humean mosaic together if not spacetime relations? How do laws `govern' or `produce' physical states or facts in a non-temporal setting? In the following section, we aim to identify and articulate central features of the QG approaches discussed here that may contribute to address these fundamental questions.   

\section{Quantum gravity seeds of a Humean mosaic}\label{entanglement}

In this section, we want to highlight the key constitutive role played by quantum entanglement in the QG formalisms based on spin networks, spin foams and discrete quantum geometry, which were introduced in the previous section. We will thus contribute to further strengthen the suggestion that quantum entanglement is a fundamental feature at the QG level and a crucial ingredient in trying to make sense of laws at this level. The fundamental role of quantum entanglement in QG, and in the context of the emergence of spacetime, has been hinted at on the basis on various different theoretical results, such as the Ryu-Takayanagi entropy formula (\cite{VanRaamsdonk, Faulkner,Qi}; see \cite{Jaksland:2021aa} and \cite{Cinti:2021aa} for philosophical discussions). We further argue that quantum entanglement lies at the heart of the main quantum gravity formalisms considered here, in a way that naturally suggests an entanglement-based non-spatio-temporal Humean basis for the quantum gravity laws (that is, taking quantum entanglement as a fundamental `gluing relation' among non-spatio-temporal \lq facts\rq).   

Let us first consider the fact that one can relate the topological and quantum geometric properties encoded in spin network states to the entanglement of their constitutive algebraic degrees of freedom (\cite{BianchiLivine,ColafranceschiOriti}). Indeed, the combinatorics of the graph underlying spin network states can be obtained as an entanglement pattern characterizing the quantum correlations of the degrees of freedom of the spin network vertices (the quantum 3-simplices in the simplicial geometry picture) (\cite{BaytasBianchi,ColafranceschiOriti}). A spin network state associated to a (closed) graph $\gamma$ is indeed obtained from product states of (disconnected) spin network vertices---each formed by tensoring the basis elements in the Hilbert space of a single spin network vertex or quantized simplex (or, more generally, quantized polyhedron)---by the imposition of a `gluing projector' enforcing maximal entanglement across the quantum degrees of freedom (representation spaces) associated to pairs of would-be shared faces (boundary triangles, in the 4d simplicial picture, in which one deals with 3-simplices). This can be done via a projector:

\begin{equation}
P^{x\otimes y}_{i} : \mathcal{H}^x_i\bigotimes \mathcal{H}^y_i \rightarrow Inv(\mathcal{H}^x_i\bigotimes \mathcal{H}^y_i) \quad ,
\label{entanglement-projector}
\end{equation}
that takes the product state of two spin network vertices (tetrahedra) $x$ and $y$ onto the subspace of states invariant under the diagonal action of $SU(2)$ on the two representation spaces associated to two semilinks (which are glued to form the link $i$ of the graph), by simply tracing over the corresponding $SU(2)$ labels.

\begin{figure}[!ht]
\centering
\includegraphics[width=0.6\linewidth]{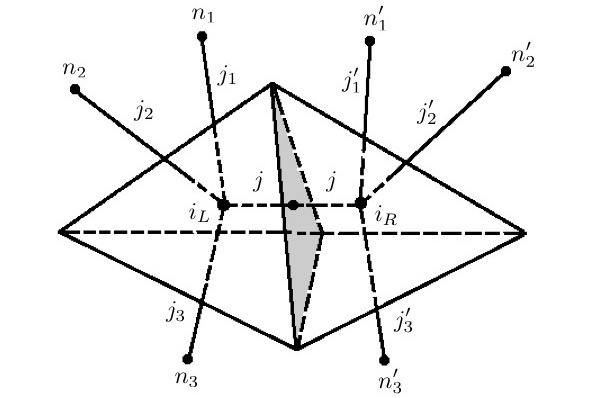}
\caption{The gluing of two spin network vertices (equivalently, two quantum tetrahedra) obtained by the imposition of maximal entanglement across the shared semi-link degrees of freedom (equivalently, the quantum data associated to a triangle on their boundary).}
\end{figure}

The adjacency matrix of the graph one intends to build is used to define a `gluing operation' (composed of all the necessary pairwise entanglement projectors) turning a product state of $V$ quanta of `space' into a spin network state for a graph of $V$ vertices (and the given adjacency matrix):

\begin{equation}
| \psi_\gamma\rangle = \prod_{A^i_{xy} = 1} P^{x\otimes y}_{i}     | \psi\rangle
\end{equation}

where the product is taken over all the non-zero entries of the adjacency matrix, thus every link $i$ connecting every pair $xy$ of vertices.
  
The construction is valid in all the quantum gravity formalisms using spin networks as fundamental quantum structures, and it is especially natural in the tensorial group field theory formalism in which the fundamental entities are organized at first in a Fock space of spin network vertices/quanta, taken as fundamental building blocks of space. To the extent that the topology of space is to be reconstructed from the combinatorial structure of (the superposition of) spin network states---this is definitely the case in spin foam models and tensorial group field theories, and arguably so in canonical quantum gravity as well---then it seems natural to conclude that the spatial topology is grounded on entanglement relations.

What about (quantum) geometry (\cite{ColafranceschiOriti})? First of all, we have seen that the gluing of two quantum tetrahedra (or two semi-links incident to the spin network vertices dual to them) across a shared triangle (to form a link connecting the two spin network vertices) is achieved by imposing a maximal entanglement relation between the states associated to the two quantum tetrahedra, specifically between their components associated to the shared triangle. The simplest measure of such `link entanglement' or `triangle entanglement' is given by––is scaling with––the number of entangled degrees of freedom, in turn given by the dimension of the Hilbert space of entangled states, i.e. the Hilbert space of the triangle now shared across the tetrahedra; this is $D = 2 j + 1$, where $j$ is the representation label (spin) associated to the triangle (or to the dual link). In the corresponding (simplicial) quantum geometry, as read by the resulting spin network state, this is either itself the area of the triangle (in a given choice of quantization map/operator ordering) or scaling just like the area.

Along the same line, it is then natural to investigate whether other geometric quantities can be similarly related to entanglement measures for spin network degrees of freedom. A simple one is the volume of the quantum tetrahedra corresponding to spin network vertices. The quantum states associated to these vertices can themselves be understood as the result of taking basis states in the tensor product of the four Hilbert spaces associated to the four triangles bounding each tetrahedron and imposing their maximal entanglement. In other words, the quantum states of a tetrahedron are the maximally entangled states that can be found in this fourfold tensor product. The maximal entanglement projection coincides with the projection onto gauge invariant states under the action of $SU(2)$, such that the states so obtained correspond to invariant tensors, i.e. linear combinations of group intertwiners. Again, the simplest measure of entanglement  for the resulting states is given by the dimension of their Hilbert space, scaling with the intertwiner label. The same quantum number labels also the spectrum of the volume operator for the tetrahedron (with the precise form of the spectrum depending on the quantization map, and other quantization choices), and the two quantities scale analogously.

The relations between entanglement in spin network states and quantum geometric quantities extend beyond these simple examples, and the above considerations actually lay the ground for a number of interesting results, also by using the correspondence between quantum gravity states and generalised tensor networks (\cite{Chirco,ChircoColafranceschi,ColafranceschiLangescheidt,Han,Singh}). For us, they can be understood as a preliminary support for the suggestion that quantum gravity may lead to a radically different reformulation of the Humean supervenience basis for the laws of nature, a purely quantum and non-spatiotemporal one, where non-spatiotemporal algebraic degrees of freedom are held together by quantum correlations (entanglement), which in turn ground the emergence of spacetime topology and geometry. Let us stress how this would go much beyond the suggestion to simply {\it add} quantum correlations to the Humean mosaic of spacetime-localised facts, since also the latter would be reduced to the former.

\section{Quantum gravity seeds for nomic production}\label{transition}

One of the greatest difficulties at the heart of the QG endeavour is the specification of a generalized, not explicitly (spatio-)temporal notion of dynamics. As noted in \cite{Lam:2023aa}, the latter––if specifiable at all––would also help grounding a non-(spatio-)temporal notion of nomic production, which in turn would allow to articulate revised non-Humean conceptions of laws (such as primitivism and dispositionalism) in a QG context without spacetime. In this section, we aim to highlight fundamental features of QG dynamics that could constitute the seeds, on the philosophical side, of non-(spatio-)temporal nomic production, and, on the physics side, of the `proto-causal' structures that could lead to the emergence of the continuum notions of time and causality. We will see that these seeds also possess distinctive quantum information-theoretic aspects.

More specifically, we focus on the quantum gravity transition amplitudes (\ref{spin foam model}) for (linear superpositions of) spin network states (as well as for more general observables), in the canonical loop quantum gravity, spin foam models or group field theory formalisms. In particular, we need to identify some notion of `ordering' among their arguments, and/or among their constitutive elements, which is crucial for any causal interpretation and any basis for nomic production. These transition amplitudes must not be invariant under switch of their arguments, and instead such switch should be corresponding to some discrete transformation, for example (taking onboard the insights of causal propagators in quantum theory) a complex conjugation:
\begin{equation}
\langle \Phi_\gamma(j,i) | \Psi_{\gamma'}(j',i') \rangle = \overline{\langle \Psi_\gamma'(j',i) | \Phi_{\gamma}(j,i) \rangle}
\label{spinfoam-ordering}
\end{equation}





Quantum gravity `processes' or `dynamics'––again, in a non-(spatio-)temporal sense––as encoded in spin foam models and lattice gravity path integrals are based on oriented 2-complexes dual to oriented cellular complexes of topological dimension 4 (for 4d gravity models). As we have seen, the 1-skeleton of the spin foam 2-complex, made of `interaction' vertices and  `propagation' links, is thus a directed graph. For Lorentzian models (whose quantum geometric data admit an interpretation in terms of Lorentzian discrete geometries), one could give a causal interpretation to the oriented links, in terms of elementary causal relations between the ordered vertices/events they connect (`causal' is here understood in the sense of some notion of `ordering', playing the role of some `proto-causality', seed from which causal relations as we know them would emerge in some appropriate continuum and classical approximation) (\cite{MarkopoulouSmolin,LivineOriti}). The assignment of quantum data to this directed graph amounts to the association of a Hilbert space $\mathcal{H}_v$ on each link (the Hilbert space of a spin network vertex or tetrahedron), so that one can then consider tensor products for unordered (not causally related) links, and of elementary `evolution' operators $\widehat{V}_a$ to the vertices, mapping the (tensor product of) Hilbert spaces of the incoming links to the (tensor product of) Hilbert spaces of the outgoing links. One can also include `gluing' operators $\widehat{K}_e$ on the links, mapping each link Hilbert space to itself (upon some appropriate dualization; this can also be obtained by associating to each oriented link two copies of the link Hilbert space, one `on the side of each vertex connected by the link'). Different choices for such operators and their kernels $\mathcal{V}_a$ and $\mathcal{K}_e$ correspond to different quantum gravity models one can considers within the same formalism (covariant counterpart of canonical loop quantum gravity, spin foam models, tensorial group field theories or lattice quantum gravity).

\begin{figure}[!ht]
\centering
\includegraphics[width=0.8\linewidth]{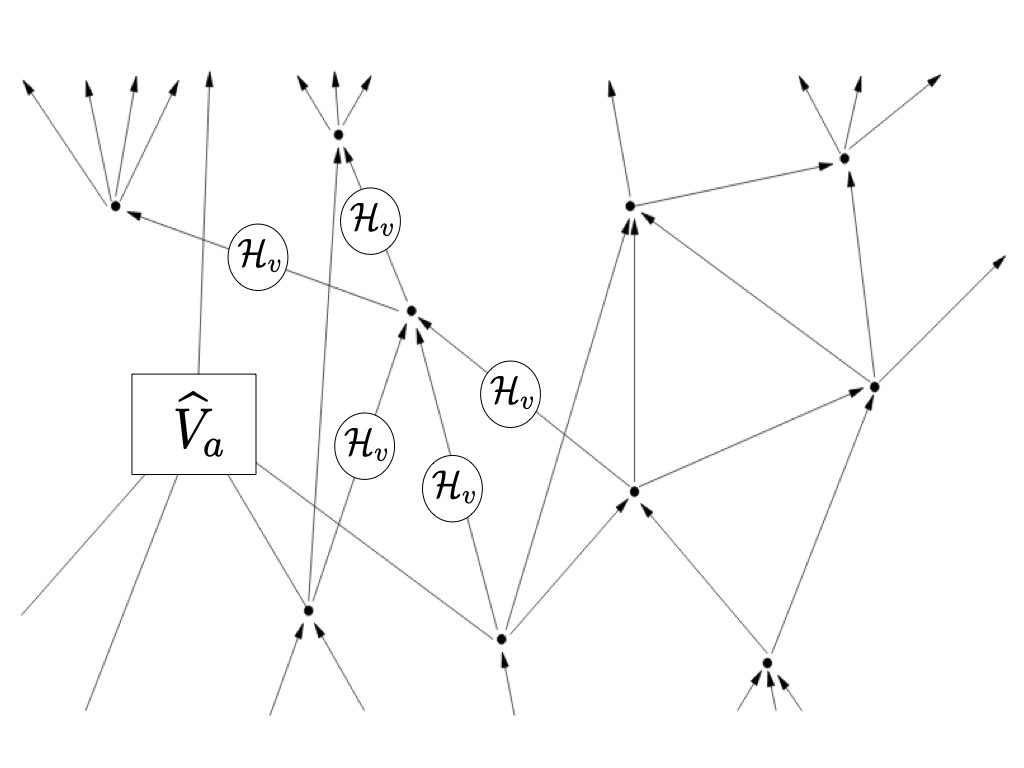}
\caption{A directed graph (dual to a simplicial 4-complex) with an assignment of quantum Hilbert spaces and elementary evolution operators, thus, potentially, a quantum causal history.}
\end{figure}

This assignment of operators allows to construct quantum `transition' amplitudes associated to the directed graph (and the 2-complex and the dual cellular complex) obtained by contracting the corresponding operator kernels across the shared Hilbert spaces (tracing over a complete basis of states): 

\begin{equation}
\mathcal{A}_\Gamma = Tr_{\{\mathcal{H}\}}\left( \prod_{e} \mathcal{K}_{e\in\Gamma}\,\prod_{a\in\Gamma}\mathcal{V}_{a}\right) \quad .
\label{QCH-amplitude}
\end{equation}

The resulting amplitude gives the spin foam amplitude associated to the directed graph, or the spin foam 2-complex, or the lattice path integral on the dual cellular complex (depending on the variables used to express it, i.e. the basis in which the contractions are computed); these are the amplitudes we had encountered in section \ref{structures}. The amplitudes for a single directed graph/complex could (should?) be then summed over to give the total transition amplitude (\ref{spin foam model}) between boundary (`initial-final') data.

Now, depending on the specific properties of the directed graph, interpreted causally, and of the operator assignment, each process described by it can be seen as defined by a {\it quantum causal history}
(\cite{Markopoulou,HawkinsMarkopoulou}), and in turn quantum causal histories can be represented as quantum computational networks (\cite{LivineTerno}). 

Such quantum causal histories encoding transition amplitudes could be interpreted in terms of some form of nomic production that is not (spatio-)temporal in any explicit sense. The quantum causal histories and the transitions amplitudes would then be understood in primitivist terms (i.e. as primitive laws) or in dispositionalist terms, encoded in the combinatorial/algebraic data of the fundamental `kinematical' structures (e.g. as represented by the relevant directed graphs). Within this non-Humean framework, these latter structures (degrees of freedom) would then constitute the ontological ground that gives rise (necessarily!), through the action of primitive laws or fundamental dispositions encoded in the transition amplitudes, to the totality of the quantum gravity facts (degrees of freedom)––the whole being possibly represented as (a superposition of) quantum causal histories. However, this general picture is connected to a number of interrelated conceptual and technical issues and difficulties that we believe are important to highlight here (without entering into too many details though).

First, for a `process' to be a quantum causal history, in the usual sense of quantum information, one needs it to correspond to a unitary evolution. This in turn requires each elementary interaction vertex operator to be unitary\footnote{There are generalizations involving completely positive maps (\cite{HawkinsMarkopoulou}) instead, and reasons to prefer them, but this does not affect the general points we want to make.} as well as a number of conditions on the evolution operator (built from the elementary ones) associated to the transition between any two acausal subsets. In fact, beyond conditions on the operators, one also needs a combinatorial condition on the directed graph: it cannot just be a directed graph, but it has to be an acyclic directed graph, i.e. a partially ordered set. In the causal interpretation of the ordering (orientation of the links), it has to be a causal set: there should not be closed causal loops of events (see also below). To the extent in which these conditions are satisfied, one also has a quantum computational network (\cite{LivineTerno}). For other constructions in the same quantum information spirit, and also motivated by (as well as with many implications for) quantum gravity, see (\cite{ArrighiMartiel}) and all the work on indefinite causal structures (\cite{OreshkovCostaBrukner}).

Second, the quantum gravity dynamics will most likely involve a superposition of the above `proto-causal' processes. One can then ask the conditions defining quantum causal histories to apply to the `final' amplitudes obtained by such superposition, and analyse what needs to be the relation between the properties of the `evolution' operators of the individual processes and those of the complete ones (\cite{LivineOriti}). One upshot of the analysis is that a superposition of quantum causal histories is not a quantum causal history. Notice that an analogous tricky relation between unitarity at the microscopic level of possible individual quantum processes and unitarity of any effective macroscopic evolution arise when the latter is obtained through some coarse graining procedure. The choice is then at which level one should expect (or impose) unitary and causal evolution, in the sense of standard quantum mechanics. Quantum gravity may well not be unitary in the general and most fundamental regime, but this is a very concrete example of the tension between the expected quantum gravity setting and the requirements from standard quantum theory and quantum information.

Third, the directed graphs corresponding to quantum gravity processes, in the frameworks singled out above, and dual to oriented cellular complexes will not be restricted, in general, to correspond to globally hyperbolic geometries (nor, in fact, discrete manifolds, containing in general various kinds of topological singularities, unless selected by hand to be regular). Non-trivial topologies, in classical continuum general relativity, imply either degenerate geometries (in the best case, at isolated points) or closed timelike curves. One could then worry about these possible obstructions to standard unitary information processing, even for given directed graph. Indeed, causal loops are easily found in the directed graphs used in quantum gravity formalisms (degenerate geometric configurations too, in fact), so they are not, in general, causal sets. Faced with this situation, one can consider three possible strategies: a) define quantum amplitudes (thus models with quantum dynamics) that suppress (if not exclude entirely) closed causal loops; b) make sure that such closed causal loops are harmless, in the sense of quantum information processing; the conditions required for a causal loop to be harmless have been discussed, for example, in \cite{LivineTerno}; c) make sure that, even if present in individual quantum processes, their quantum superposition (or coarse graining) gives a non-pathological evolution for observables, at least in an effective macroscopic regime (so that causal pathologies are confined to unobservables phenomena). It must be clear that the non-(spatio-)temporal nomic production interpretation of these transition amplitudes is only available within strategy a) suppressing closed causal loops. Indeed, as \cite[ch. 6]{Maudlin2007} makes very clear in the context of primitivism about laws, the very notion of (nomic) production is in tension (incompatible) with closed causal loops, since it would undermine the sort of productive explanation that is at the heart of this non-Humean account (the same holds true for dispositionalism).     

Fourth, there is another way in which quantum gravity amplitudes encoding its fundamental dynamics can fail to describe a standard causal evolution; this is related to the issue of unitarity above, but not identical. They may be `causally indifferent' by always summing symmetrically over opposite spacetime orientations, that is, in our interpretation, opposite causal directions. The fact that covariant quantum gravity amplitudes can be symmetric under spacetime orientation reversal is no mystery and it has been elucidated already in the classical works on the gravitational path integral (\cite{HalliwellHartle}). It is what one should expect (in fact, require) when these amplitudes define the canonical inner product between quantum gravity states, solving the canonical counterpart of (quantum) diffeomorphism invariance constraints (in fact, a slightly larger symmetry than diffeomorphism invariance in the Lagrangian formalism). Amplitudes invariant under Lagrangian diffeomorphisms but registering the spacetime orientation, and transforming into their complex conjugate under reversal, as in (\ref{spinfoam-ordering}), can also be constructed in both covariant and canonical language, and would correspond to the quantum gravity analogue of the time-ordered 2-point functions for particles and fields (the ordering would not depend on any specific time coordinate in the quantum gravity case, of course). How the orientation of the underlying complex is registered in the spin foam amplitudes and lattice gravity path integral has been discussed in \cite{LivineOriti}. All the most developed spin foam (and tensorial group field theory) models are, it turns out, orientation-symmetric; as such, they may indeed correspond to a definition of the canonical inner product (modulo the issue of non-trivial topologies) and do not correspond to quantum causal histories, neither at the level of individual processes not at the level of the complete amplitudes. It must be clear that these symmetry features for amplitudes would constitute a difficulty for any interpretation in terms of (nomic) production, since the latter seems to irreducibly involve some asymmetry. As a consequence, the non-Humean nomic production conceptions seem to fundamentally require some sort of orientation (direction), from the entities (e.g. degrees of freedom) `producing' to those `being produced'. The extent to which this is the case in the QG contexts considered here remains very much an open issue: indeed, models possessing orientation-dependent, thus possibly properly causal amplitudes can be constructed, however, as straightforward modifications of orientation-symmetric spin foam (and tensorial group field theory) models, albeit in a rather ad hoc manner (\cite{LivineOriti, Engle, BianchiDussaud}).

\section{An alternative perspective: an epistemic view on the world and its laws}\label{epsitemic}
So far we have adopted what could be called an ontology-first perspective. The questions we addressed have been what quantum gravity structures are suggested as fundamental in modern QG formalisms and how they can ground laws of nature. The latter are themselves interpreted from such an ontology-first (realist) perspective, as being somehow `out there' in the world, and then eventually, discovered by epistemic agents. They may be ontological primitives or supervenient, but they are indeed out there and perspective-independent, objective. 

In many ways, this is the traditional and most worked-out approach to laws of nature. It is also the immediate, default perspective of most quantum gravity scientists, and in fact, of most scientists in any discipline: according to this intuitive view, scientists are after the objective description of what is out there and of the laws governing it, from a detached, third-person point of view.

But it is important to note here that this is not the only possible or reasonable approach to the issues we focused on in this contribution. In many ways, the perspective- or context-independent view of reality is challenged in the quantum gravity formalisms we have considered and, more generally, in (certain approaches to) quantum theory itself; these challenges may affect not only the philosophical issue of the nature of laws, but also the very developments of physical theorizing in the QG context. 

First, central in much quantum gravity literature, including the formalisms we focused on, is a relational perspective on classical and quantum spacetime and gravity. According to this perspective, the observables in terms of which the physics of spacetime, geometry, gravity and matter is encoded are not associated to the manifold on which dynamical fields are initially defined, but only to manifold-independent relational among their possible values. In particular, any localization in space or time, and thus any meaningful notion of temporal evolution, has to be defined with respect to some physical reference frame constructed via physical and dynamical fields chosen as clock and rods. If fields are taken as fundamental entities, and diffeomorphism invariance is to be maintained, then this relational construction is to also implemented in the full quantum theory. If fields, spacetime and geometry are emergent notions, then this relational understanding is to be recovered in some appropriate approximate regime of the non-spatiotemporal theory.  The point remains that one can only have, in general (i.e. beyond special solutions or asymptotic boundary conditions, etc.), only a relational, context-dependent (in the sense of dependent on chosen physical frame) definition of space and time, and no guarantee of invariance of physical predictions under change of physical reference frames (recall that we are not talking of coordinate or idealized reference frames here, to which relativistic general covariance refers).

Second, in the classical context, one can argue that this context-dependence remains `objectified', i.e. that one has an objective description of what relational properties defining space and time are, that spacetime properties, although only defined with respect to some physical reference frame, are objectively so defined. The ontology of the world may be relational in this sense, but not less solid and independent of epistemic agents writing down the mathematical description of such relational ontology and the laws governing it (in other words, an appropriate relational/structural version of scientific realism fully applies to spacetime, such as ontic structural realism about spacetime, see \cite{Esfeld:2008aa}). At most, if epistemic agents have any role at all, they can be themselves fully accounted for in their embodiment as physical reference frames. 

This epistemically somewhat insensitive perspective can be challenged at the quantum level. In quantum theory, it is much less clear that properties of physical systems, relational or not, are observer independent, defined independently of the epistemic agents using the quantum mechanical models in which they are formulated. We refer here to the broad class of epistemic interpretations of quantum mechanics, and of quantum states in particular, i.e. all the interpretations in which quantum states are not ontic but a summary of the knowledge, past interaction history, or beliefs of the epistemic agent using the quantum formalism to model the systems they refer to, depending on the specific brand of epistemic interpretation. 
To the extent in which quantum gravity is framed within the standard quantum formalism, as it is the case in the quantum gravity approaches we focused on, this epsitemic perspective could then be applied to the non-spatiotemporal structures we discussed. They would then be relational in a stronger sense than in classical GR, and the sense in which they are physically real would be even more strongly perspectival.  

Third, one can take an epistemic perspective on laws of nature too. That is, one can deny that laws of nature are `out there' and maintain instead that they are entirely in the epistemic agents' minds, and correspond to specific elements in our models of the world. 
This view can be argued for on purely philosophical grounds (as a form of anti-realism about laws for instance), but it clearly resonates with the points recalled above concerning GR and quantum theory. 
From this viewpoint, it is objected to the standard Humean and non-Humean realist conceptions of laws we have considered in this contribution that they all `illegitimately' reify what are actually epistemic constructions in the first place (and what should be considered as such)––and such reification then leads to tension with developments in physics that do not conform such reified pictures, as in the context of QG where spatio-temporal notions may not be available. On the contrary, in a more epistemic perspective, the QG developments pointing towards the disappearance of spacetime at some fundamental level are less `problematic' in a sense and can be naturally accounted for in purely epistemic terms.       
Quantum gravity will remain peculiar, with respect to more standard physical theories, but its `laws' will not be of a radically different kind. One would maintain that a quantum gravity theory is the result of the construction work of epistemic agents, and like any other such theoretical construction it 
is a successful explanatory scheme for our interactions with the world. One peculiarity would be that it will remain for the most part not directly related to experiences (but fundamental physics models rarely are), and thus most of its statements will be 
grasped by analogy and through their role within the theoretical scheme, more so than what happens already for other theories of modern physics. From this point of view, one would have to seriously face 
the usual philosophical challenges of elucidating the precise role, impact and limitations of epistemological tools like analogy, abstractions, idealization, counterfactual reasoning, and so on. The QG context will be even more mathematically abstract and conceptually tricky than usual, and the ground for counterfactuals as well as hypothetical statements will have to be found only in their explanatory power and their (probably very indirect) empirical support, i.e. in epistemological rather than ontological considerations. 

Finally, a comment on the role of information-theoretic tools and concepts, which have started to play a prominent role in quantum gravity. 
In the latter context, we have no spacetime notions to rely on at a more fundamental level, and thus we have to theorize and model without spacetime. But without spacetime and geometry, we are left with data expressed in combinatorial and algebraic terms, as well as with information processing rules for such data. Quantum computing then provides abstract models of quantum information processing––and to a certain extent of our own reasoning (at least when it comes to classical computing). The use of quantum information processing tools then follows rather naturally from this––again, purely epistemic––perspective.

\section{Conclusions}\label{conclusion}


Traditional accounts of laws of nature face serious challenges in a context in which spacetime is not fundamental, as suggested by modern developments towards a theory of quantum gravity and quantum spacetime. In this paper, we have looked at some of these challenges with a specific focus on quantum gravity formalisms sharing similar structures both at the level of quantum states and of dynamical processes, both of a purely combinatorial and algebraic kind. We have identified possible seeds for what could represent, at a philosophical level, a novel non-spatiotemporal ground for both Humeanist (QG non-spatio-temporal `glue') and primitivist / dispositional perspectives (QG non-temporal nomic production) on laws. These seeds are also key elements for the eventual reconstruction of spacetime, geometry and causality in some approximation. Remarkably, these seeds are of a distinctively quantum information-theoretic nature, being based on the role played by entanglement in quantum gravity states and on the possible reformulation of quantum gravity processes as quantum causal histories (or quantum circuits). This confirms the fertile overlap between quantum gravity and quantum information theory. We have also discussed, albeit briefly, how the same issues appear from an epistemic perspective on laws of nature, which represents another interesting direction for further research.

\section*{Acknowledgments}
The authors thank the editor of this special issue for his remarkable patience. We thank E. Margoni, C. W\"{u}thrich for discussions. This work has been supported in part by grants from DFG and FQxI (to DO). VL is grateful to the Swiss National Science Foundation for financial support (PP00P1\_211010).

\printbibliography

@article{lamwut18,
	Author = {Lam, Vincent and W\"uthrich, Christian},
	Journal = {Studies in History and Philosophy of Modern Physics},
	Pages = {39-51},
	Title = {Spacetime is as spacetime does},
	Volume = {64},
	Year = {2018}}

@article{MargoniOriti,
	Author = {Margoni, Emilia and Oriti, Daniele},
	Journal = {},
	Title = {Spacetime emergence in quantum gravity: what role for functionalism?},
    Year = {2024},
    eprint = "2404.11386",
    archivePrefix = "arXiv",
    primaryClass = "physics.hist-ph",
}

@article{OritiDisEmerg,
    author = "Oriti, Daniele",
    title = "{Disappearance and emergence of space and time in quantum gravity}",
    eprint = "1302.2849",
    archivePrefix = "arXiv",
    primaryClass = "physics.hist-ph",
    reportNumber = "AEI-2013-050",
    doi = "10.1016/j.shpsb.2013.10.006",
    journal = "Stud. Hist. Phil. Sci. B",
    volume = "46",
    pages = "186--199",
    year = "2014"
}

@book{Huggett-Wuthrichforth,
	Author = {Huggett, N. and W{\"u}thrich, C. },
	Publisher = {Oxford: Oxford University Press},
	Title = {Out of Nowhere},
	Year = {forthcoming}}

@article{OritiLevels,
    author = "Oriti, Daniele",
    title = "{Levels of spacetime emergence in quantum gravity}",
    eprint = "1807.04875",
    archivePrefix = "arXiv",
    primaryClass = "physics.hist-ph",
    month = "7",
    year = "2018"
}

@article{Crowther,
    author = "Crowther, Karen",
    title = "{As below, so before: \textquoteleft{}synchronic\textquoteright{} and \textquoteleft{}diachronic\textquoteright{} conceptions of spacetime emergence}",
    eprint = "1912.12065",
    archivePrefix = "arXiv",
    primaryClass = "physics.hist-ph",
    doi = "10.1007/s11229-019-02521-1",
    journal = "Synthese",
    volume = "198",
    number = "8",
    pages = "7279--7307",
    year = "2021"
}

@inproceedings{Seiberg,
    author = "Seiberg, Nathan",
    title = "{Emergent spacetime}",
    booktitle = "{23rd Solvay Conference in Physics: The Quantum Structure of Space and Time}",
    eprint = "hep-th/0601234",
    archivePrefix = "arXiv",
    doi = "10.1142/9789812706768_0005",
    pages = "163--178",
    month = "1",
    year = "2006"
}

@inbook{Padmanabhan,
    author = "Padmanabhan, Thanu",
    editor = "Ashtekar, Abhay and Petkov, Vesselin",
    title = "{Gravity and Spacetime: An Emergent Perspective}",
    booktitle = "{Springer Handbook of Spacetime}",
    doi = "10.1007/978-3-642-41992-8_12",
    pages = "213--242",
    year = "2014"
}

@book{OritiApproaches,
    author = "Oriti, Daniele",
    title = "{Approaches to quantum gravity: Toward a new understanding of space, time and matter}",
    isbn = "978-0-521-86045-1, 978-0-511-51240-7",
    publisher = "Cambridge University Press",
    month = "3",
    year = "2009"
}

@book{Handbook,
    editor = "Bambi, Cosimo and Modesto, Leonardo and Shapiro, Ilya",
    title = "{Handbook of Quantum Gravity}",
    doi = "10.1007/978-981-19-3079-9",
    isbn = "978-981-19307-9-9",
    publisher = "Springer",
    year = "2023"
}

@article{Darby2012,
	Author = {Darby, George},
	Journal = {British Journal for the Philosophy of Science},
	Pages = {773-788},
	Title = {Relational holism and {Humean} supervenience},
	Volume = {63},
	Year = {2012}}

@book{Maudlin2007,
	Address = {Oxford},
	Author = {Maudlin, Tim},
	Publisher = {Oxford University Press},
	Title = {The Metaphysics Within Physics},
	Year = {2007}}

@article{LQG,
    author = "Giesel, Kristina and Sahlmann, Hanno",
    editor = "Barrett, John and Giesel, Kristina and Hellmann, Frank and Jonke, Larisa and Krajewski, Thomas and Lewandowski, Jerzy and Rovelli, Carlo and Sahlmann, Hanno and Steinacker, Harold",
    title = "{From Classical To Quantum Gravity: Introduction to Loop Quantum Gravity}",
    eprint = "1203.2733",
    archivePrefix = "arXiv",
    primaryClass = "gr-qc",
    reportNumber = "APCTP-PRE2012-004, APCTP Pre2012-004",
    doi = "10.22323/1.140.0002",
    journal = "PoS",
    volume = "QGQGS2011",
    pages = "002",
    year = "2011"
}

@article{SF,
    author = "Perez, Alejandro",
    title = "{The Spin Foam Approach to Quantum Gravity}",
    eprint = "1205.2019",
    archivePrefix = "arXiv",
    primaryClass = "gr-qc",
    doi = "10.12942/lrr-2013-3",
    journal = "Living Rev. Rel.",
    volume = "16",
    pages = "3",
    year = "2013"
}

@inproceedings{GFT,
    author = "Oriti, Daniele",
    title = "{The microscopic dynamics of quantum space as a group field theory}",
    booktitle = "{Foundations of Space and Time: Reflections on Quantum Gravity}",
    eprint = "1110.5606",
    archivePrefix = "arXiv",
    primaryClass = "hep-th",
    pages = "257--320",
    month = "10",
    year = "2011"
}

@article{Lam:2023aa,
	Author = {Lam, Vincent and W{\"u}thrich, Christian},
	Da = {2023/08/24},
	Doi = {10.1007/s11229-023-04305-0},
	Id = {Lam2023},
	Isbn = {1573-0964},
	Journal = {Synthese},
	Number = {3},
	Pages = {71},
	Title = {Laws beyond spacetime},
	Ty = {JOUR},
	Url = {https://doi.org/10.1007/s11229-023-04305-0},
	Volume = {202},
	Year = {2023}}

@article{VanRaamsdonk,
    author = "Van Raamsdonk, Mark",
    title = "{Building up spacetime with quantum entanglement II: It from BC-bit}",
    eprint = "1809.01197",
    archivePrefix = "arXiv",
    primaryClass = "hep-th",
    month = "9",
    year = "2018"
}

@inproceedings{Faulkner,
    author = "Faulkner, Thomas and Hartman, Thomas and Headrick, Matthew and Rangamani, Mukund and Swingle, Brian",
    title = "{Snowmass white paper: Quantum information in quantum field theory and quantum gravity}",
    booktitle = "{Snowmass 2021}",
    eprint = "2203.07117",
    archivePrefix = "arXiv",
    primaryClass = "hep-th",
    reportNumber = "BRX-TH-6703",
    month = "3",
    year = "2022"
}

@article{Qi,
    author = "Qi, Xiao-Liang",
    title = "{Does gravity come from quantum information?}",
    doi = "10.1038/s41567-018-0297-3",
    journal = "Nature Phys.",
    volume = "14",
    number = "10",
    pages = "984--987",
    year = "2018"
}

@article{Jaksland:2021aa,
	Author = {Jaksland, Rasmus},
	Da = {2021/10/01},
	Doi = {10.1007/s11229-020-02671-7},
	Id = {Jaksland2021},
	Isbn = {1573-0964},
	Journal = {Synthese},
	Number = {10},
	Pages = {9661--9693},
	Title = {Entanglement as the world-making relation: distance from entanglement},
	Ty = {JOUR},
	Url = {https://doi.org/10.1007/s11229-020-02671-7},
	Volume = {198},
	Year = {2021}}

@article{Cinti:2021aa,
	Author = {Cinti, Enrico and Sanchioni, Marco},
	Da = {2021/12/01},
	Doi = {10.1007/s11229-021-03270-w},
	Id = {Cinti2021},
	Isbn = {1573-0964},
	Journal = {Synthese},
	Number = {3},
	Pages = {10839--10863},
	Title = {Humeanism in light of quantum gravity},
	Ty = {JOUR},
	Url = {https://doi.org/10.1007/s11229-021-03270-w},
	Volume = {199},
	Year = {2021}}

@article{ColafranceschiOriti,
    author = "Colafranceschi, Eugenia and Oriti, Daniele",
    title = "{Quantum gravity states, entanglement graphs and second-quantized tensor networks}",
    eprint = "2012.12622",
    archivePrefix = "arXiv",
    primaryClass = "hep-th",
    doi = "10.1007/JHEP07(2021)052",
    journal = "JHEP",
    volume = "07",
    pages = "052",
    year = "2021"
}

@inbook{BianchiLivine,
    author = "Bianchi, Eugenio and Livine, Etera R.",
    title = "{Loop Quantum Gravity and Quantum Information}",
    eprint = "2302.05922",
    archivePrefix = "arXiv",
    primaryClass = "gr-qc",
    doi = "10.1007/978-981-19-3079-9_108-1",
    year = "2023"
}

@article{BaytasBianchi,
    author = "Bayta\c{s}, Bekir and Bianchi, Eugenio and Yokomizo, Nelson",
    title = "{Gluing polyhedra with entanglement in loop quantum gravity}",
    eprint = "1805.05856",
    archivePrefix = "arXiv",
    primaryClass = "gr-qc",
    reportNumber = "IGC-18-5-1",
    doi = "10.1103/PhysRevD.98.026001",
    journal = "Phys. Rev. D",
    volume = "98",
    number = "2",
    pages = "026001",
    year = "2018"
}

@article{Chirco,
    author = "Chirco, Goffredo",
    title = "{Holographic Entanglement in Group Field Theory}",
    doi = "10.3390/universe5100211",
    journal = "Universe",
    volume = "5",
    number = "10",
    pages = "211",
    year = "2019"
}

@article{ChircoColafranceschi,
    author = "Chirco, Goffredo and Colafranceschi, Eugenia and Oriti, Daniele",
    title = "{Bulk area law for boundary entanglement in spin network states: Entropy corrections and horizon-like regions from volume correlations}",
    eprint = "2110.15166",
    archivePrefix = "arXiv",
    primaryClass = "hep-th",
    doi = "10.1103/PhysRevD.105.046018",
    journal = "Phys. Rev. D",
    volume = "105",
    number = "4",
    pages = "046018",
    year = "2022"
}

@article{ColafranceschiLangescheidt,
    author = "Colafranceschi, Eugenia and Langenscheidt, Simon and Oriti, Daniele",
    title = "{Holographic properties of superposed quantum geometries}",
    eprint = "2207.07625",
    archivePrefix = "arXiv",
    primaryClass = "quant-ph",
    month = "7",
    year = "2022"
}

@article{Han,
    author = "Han, Muxin and Hung, Ling-Yan",
    title = "{Loop Quantum Gravity, Exact Holographic Mapping, and Holographic Entanglement Entropy}",
    eprint = "1610.02134",
    archivePrefix = "arXiv",
    primaryClass = "hep-th",
    doi = "10.1103/PhysRevD.95.024011",
    journal = "Phys. Rev. D",
    volume = "95",
    number = "2",
    pages = "024011",
    year = "2017"
}

@article{Singh,
    author = "Singh, Sukhwinder and McMahon, Nathan A. and Brennen, Gavin K.",
    title = "{Holographic spin networks from tensor network states}",
    eprint = "1702.00392",
    archivePrefix = "arXiv",
    primaryClass = "cond-mat.str-el",
    doi = "10.1103/PhysRevD.97.026013",
    journal = "Phys. Rev. D",
    volume = "97",
    number = "2",
    pages = "026013",
    year = "2018"
}

@article{Esfeld:2008aa,
	Author = {Esfeld, Michael and Lam, Vincent},
	Da = {2008/01/01},
	Doi = {10.1007/s11229-006-9076-2},
	Id = {Esfeld2008},
	Isbn = {1573-0964},
	Journal = {Synthese},
	Number = {1},
	Pages = {27--46},
	Title = {Moderate structural realism about space-time},
	Ty = {JOUR},
	Url = {https://doi.org/10.1007/s11229-006-9076-2},
	Volume = {160},
	Year = {2008}}

@article{MarkopoulouSmolin,
    author = "Markopoulou, Fotini and Smolin, Lee",
    title = "{Causal evolution of spin networks}",
    eprint = "gr-qc/9702025",
    archivePrefix = "arXiv",
    reportNumber = "CGPG-97-2-1, CGPG-97-2-2A",
    doi = "10.1016/S0550-3213(97)00488-4",
    journal = "Nucl. Phys. B",
    volume = "508",
    pages = "409--430",
    year = "1997"
}

@article{HawkinsMarkopoulou,
    author = "Hawkins, Eli and Markopoulou, Fotini and Sahlmann, Hanno",
    title = "{Evolution in quantum causal histories}",
    eprint = "hep-th/0302111",
    archivePrefix = "arXiv",
    doi = "10.1088/0264-9381/20/16/320",
    journal = "Class. Quant. Grav.",
    volume = "20",
    pages = "3839",
    year = "2003"
}

@article{Markopoulou,
    author = "Markopoulou, Fotini",
    title = "{Quantum causal histories}",
    eprint = "hep-th/9904009",
    archivePrefix = "arXiv",
    reportNumber = "CGPG-99-3-4",
    doi = "10.1088/0264-9381/17/10/302",
    journal = "Class. Quant. Grav.",
    volume = "17",
    pages = "2059--2072",
    year = "2000"
}

@article{LivineOriti,
    author = "Livine, Etera R. and Oriti, Daniele",
    title = "{Implementing causality in the spin foam quantum geometry}",
    eprint = "gr-qc/0210064",
    archivePrefix = "arXiv",
    reportNumber = "DAMTP-2002-127",
    doi = "10.1016/S0550-3213(03)00378-X",
    journal = "Nucl. Phys. B",
    volume = "663",
    pages = "231--279",
    year = "2003"
}

@article{LivineTerno,
    author = "Livine, Etera R. and Terno, Daniel R.",
    title = "{Quantum causal histories in the light of quantum information}",
    eprint = "gr-qc/0611135",
    archivePrefix = "arXiv",
    doi = "10.1103/PhysRevD.75.084001",
    journal = "Phys. Rev. D",
    volume = "75",
    pages = "084001",
    year = "2007"
}

@article{ArrighiMartiel,
    author = "Arrighi, Pablo and Martiel, Simon",
    title = "{Quantum Causal Graph Dynamics}",
    eprint = "1607.06700",
    archivePrefix = "arXiv",
    primaryClass = "cs.DM",
    reportNumber = "See also the more recent https://arxiv.org/abs/2110.10587",
    doi = "10.1103/PhysRevD.96.024026",
    journal = "Phys. Rev. D",
    volume = "96",
    number = "2",
    pages = "024026",
    year = "2017"
}

@article{OreshkovCostaBrukner,
    author = "Oreshkov, Ognyan and Costa, Fabio and Brukner, Caslav",
    title = "{Quantum correlations with no causal order}",
    eprint = "1105.4464",
    archivePrefix = "arXiv",
    primaryClass = "quant-ph",
    doi = "10.1038/ncomms2076",
    journal = "Nature Commun.",
    volume = "3",
    pages = "1092",
    year = "2012"
}

@article{HalliwellHartle,
    author = "Halliwell, Jonathan J. and Hartle, James B.",
    title = "{Wave functions constructed from an invariant sum over histories satisfy constraints}",
    reportNumber = "NSF-ITP-90-97",
    doi = "10.1103/PhysRevD.43.1170",
    journal = "Phys. Rev. D",
    volume = "43",
    pages = "1170--1194",
    year = "1991"
}

@article{Engle,
    author = "Engle, Jonathan and Zipfel, Antonia",
    title = "{Lorentzian proper vertex amplitude: Classical analysis and quantum derivation}",
    eprint = "1502.04640",
    archivePrefix = "arXiv",
    primaryClass = "gr-qc",
    doi = "10.1103/PhysRevD.94.064024",
    journal = "Phys. Rev. D",
    volume = "94",
    number = "6",
    pages = "064024",
    year = "2016"
}

@article{BianchiDussaud,
    author = "Bianchi, Eugenio and Martin-Dussaud, Pierre",
    title = "{Causal structure in spin-foams}",
    eprint = "2109.00986",
    archivePrefix = "arXiv",
    primaryClass = "gr-qc",
    month = "9",
    year = "2021"
}

@book{Lewis1986,
	address = {New York},
	author = {Lewis, David},
	publisher = {Oxford University Press},
	title = {Philosophical Papers, Volume II},
	year = {1986}}

@article{Cohen:2009aa,
	Author = {Cohen, Jonathan and Callender, Craig},
	Da = {2009/07/01},
	Doi = {10.1007/s11098-009-9389-3},
	Id = {Cohen2009},
	Isbn = {1573-0883},
	Journal = {Philosophical Studies},
	Number = {1},
	Pages = {1--34},
	Title = {A better best system account of lawhood},
	Ty = {JOUR},
	Url = {https://doi.org/10.1007/s11098-009-9389-3},
	Volume = {145},
	Year = {2009}}

@book{Bird2007,
	address = {Oxford},
	author = {Bird, Alexander},
	publisher = {Oxford University Press},
	title = {Nature's Metaphysics: Laws and Properties},
	year = {2007}}

@incollection{Hall2015,
	address = {Chichester},
	author = {Hall, Ned},
	booktitle = {{A Companion to David Lewis}},
	editor = {Loewer, Barry and Schaffer, Jonathan},
	pages = {262-277},
	publisher = {Wiley},
	title = {{Humean reductionism about laws of nature}},
	year = {2015}}

@incollection{Earman1984,
	address = {Dordrecht},
	author = {Earman, John},
	booktitle = {D.\ M.\ Armstrong},
	editor = {Bogdan, Radu J},
	publisher = {D.\ Reidel Publishing Company},
	title = {Laws of nature: the empiricist challenge},
	year = {1984}}


\end{document}